\documentclass[epj,onecolumn]{svjour}

\usepackage[utf8]{inputenc} %
\usepackage{verbatim}    
\usepackage{cuted}
\usepackage{dcolumn}
\usepackage{bm}
\usepackage{epsfig}
\usepackage{epstopdf}
\usepackage{amsmath}
\usepackage{amsfonts}
\usepackage{amssymb,amscd}
\usepackage{color} 
\usepackage[normalem]{ulem} 
\usepackage{longtable} 

\newcommand{\be}{\begin{equation}}
\newcommand{\ee}{\end{equation}}
\newcommand{\ben}{\begin{eqnarray}}
\newcommand{\een}{\end{eqnarray}}

\begin{document}

\title{Heating triangle singularities in heavy ion collisions}

\author{Luciano M. Abreu\inst{1} \and Felipe J. Llanes-Estrada\inst{2}
}                    
\institute{Instituto de F\'isica, Universidade Federal da Bahia, Salvador, Bahia, 40170-115, Brazil \and 
 Dept. F\'isica Te\'orica and IPARCOS, Universidad Complutense, Madrid, 28040, Spain}

\date{April 19th 2020} 

\abstract{
We predict that triangle singularities of hadron spectroscopy can be strongly affected in heavy ion collisions.  To do it we examine various effects on the singularity-inducing triangle loop of finite temperature in the terminal hadron phase. 
It appears that peaks seen in central heavy ion collisions are more likely to be hadrons than rescattering effects under two conditions. First, the flight-time of the intermediate hadron state must be comparable to the lifetime of the equilibrated fireball (else, the reaction mostly happens {\it in vacuo} after freeze out). 
Second, the medium effect over the triangle-loop particle mass or width must be sizeable. 
When these (easily checked) conditions are met, the medium quickly reduces the singularity: at T about 150 MeV, even by two orders of magnitude, acting then as a spectroscopic filter.
}

\maketitle

\section{Introduction}
\label{Introduction}

\begin{figure}[h]
\centering
\includegraphics[width=0.5\columnwidth]{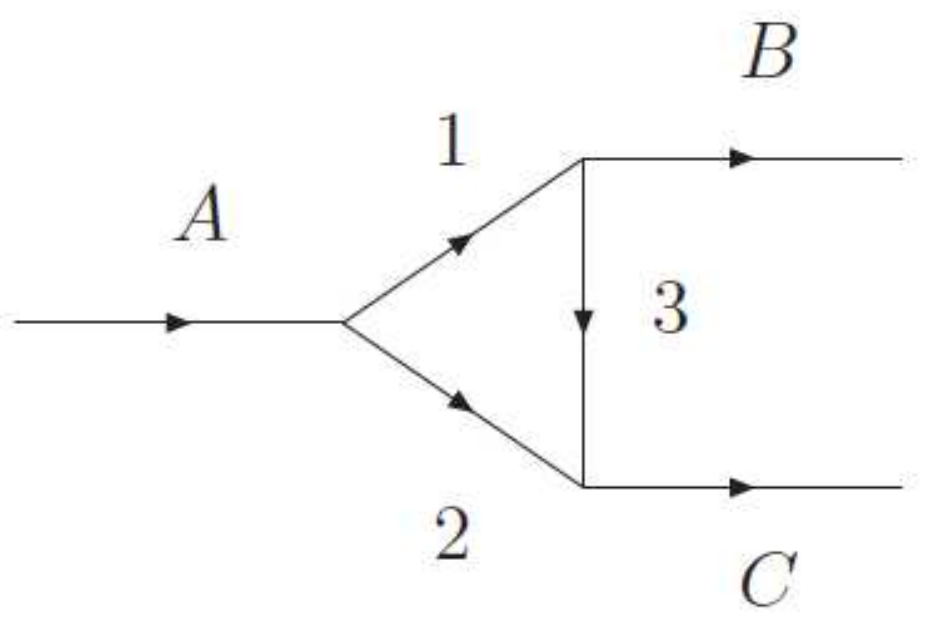} 
\caption{Triangle diagram for the reaction $A \rightarrow BC $, possibly yielding a triangle singularity~\cite{Molina:2020kyu}. }
\label{fig:triangle}
\end{figure}

At the foundation of particle physics since the 1960s is the understanding of hadrons in quark-model terms. It is thus surprising that there are so many ``structures'' in accelerator data
that remain unclassified. While there are too few baryons ($qqq$-like) in comparison to early model expectations, there are numerous claims for supernumerary meson ($q\overline{q}$) resonances. Perhaps this is the plethora of exotic resonances expected from Quantum Chromodynamics, that elevated the quark model to a field theory with sectors counting different numbers of quarks, antiquarks and gluons.  But some of those new ``hadrons''  without a clear overall pattern also beg for dynamical explanations based on how the known hadrons rescatter under their strong force. A leading candidate hypothesis to effect much of the probably needed cleanup is the concept of triangle singularities (and other cuspy features), much discussed in hadron physics in the last decade~\cite{Bugg:2011jr,Wang:2013hga,Szczepaniak:2015eza,Guo:2019twa}. Such methods are becoming standard among experimental collaborations, reexamining new and earlier ``resonance'' discoveries for singularity structures not necessarily reflecting a new particle. Serve as example the recent claim~\cite{Alexeev:2020lvq} that the $a_1(1420)$ is no axial-vector meson resonance
but such triangle singularity instead~\cite{Ketzer:2015tqa,Aceti:2016yeb}. 
There are several works that explain the mechanism~\cite{Aceti:2015zva,Xie:2016lvs,Bayar:2016ftu,Guo:2019twa,Nakamura:2019nwd,Bayar:2017svj,Molina:2020kyu}: in brief, the triangle singularity, that receives its name from the Feynman diagram in Fig.~\ref{fig:triangle} happens when the three intermediate particles become on shell and two have parallel momenta, if such kinematics is allowed.

In a different subfield, analysis of heavy-ion collision data routinely report narrow meson resonances, and an effort has developed to use them for spectroscopy~\cite{Cho:2011ew}. For example, the cross-section has been proposed to distinguish internal compositions of charmonia such as $X(3872)$ \cite{Torres:2014fxa,Abreu:2016qci,Zhang:2020dwn}, with canonical $q\overline{q}$ mesons thoroughly analyzed in the past~\cite{Brambilla:2017zei}. In contrast, molecular configurations are supposed to quickly break up in the hot medium. 

In this article we observe that dynamical singularities such as the triangle one, not associated with a new particle, can also quickly melt away, so that one can often assert that peaks in central heavy ion collision data are more likely physical hadron states. Our calculations illustrate how this singularity disappearance happens already at the lower temperature of the hadron medium, much the more so in the quark-gluon plasma phase. 
Through several examples, we examine under what conditions the phenomenon can happen, reasonably showing that this is due to the shift in the intermediate-particle pole position (due to Bose-Einstein enhancement of their decay in a light-meson populated thermal medium) erasing the kinematic coincidence that causes the singularity.

\section{Formalism: triangle loop at finite T}
\label{Formalism}

We adopt the rest frame of the decaying particle A in figure~\ref{fig:triangle}, momentum and energy conservation fix the external-variable kinematics,
\begin{eqnarray}\! \! \! \!\! \!
P^0  \equiv  E_A = m_A, \ \   
P^0 - k^0  \equiv  E_C = \frac{m_A ^2 - m_B ^2 + m_C ^2}{2 m_A},
\\ \nonumber \! \! \! \! \! \!
k^0  \equiv  E_B = \frac{m_A ^2 + m_B ^2 - m_C ^2}{2 m_A}, \ \ 
|\vec{k}| = \frac{\sqrt{\lambda(m_A ^2 , m_B ^2 , m_C ^2)} }{2 m_A},  
\end{eqnarray}
where $\lambda(a,b,c) = a^2+b^2+c^2-2(ab+ac+bc)$. 
The internal variable running in the loop is the four-momentum $q=(q^0,{\bf q})$ of particle 2, with $z \equiv \cos{\theta_q}$.
This reference frame, depending on experimental circumstances, might not coincide with that in which the thermal medium is at rest: but the additional complication of including a Lorentz boost to make the decaying particle produced in motion, as well as spin, are irrelevant to demonstrate the damping of the triangle singularity, so we take the particle as almost at rest respect to the thermal bath. This is the limit opposite to jet quenching, in which the partons are travelling much faster than the medium. 
The medium is taken to be infinite in extension and perfectly equilibrated at temperature $T$, hypothesis under which one can employ standard thermal field theory. 
The applicability of this setup to a physical heavy ion collision is briefly addressed in section~\ref{sec:flight} below. When applicable, it is of course only a first approximation, but convolving our functions with a full medium-evolution code will not bring back a singularity that has been thermally washed away.

Therefore, we consider only the scalar three-point loop integral 
\begin{eqnarray}
I_{\triangleleft} &= & i \int \frac{d^4 q}{(2\pi)^4} \frac{1}{\left(P-q)^2
-m_1^2+i\epsilon\right)} \\ \nonumber 
& & \cdot \frac{1}{
\left[(q^2-m_2^2+i \epsilon \right]\left[(P-q-k)^2-m_3^2+i \epsilon \right]}. 
\label{loop1}
\end{eqnarray}
With heavy particles in the triangle, {\it e. g.} $m_{D^{0 \pm }}\gg T$,
or as is our thrust, for singularity-specific kinematics involving decays unreachable by propagating an antiparticle, we are allowed to neglect the negative energy part of the propagators (that would otherwise have to be taken into account in a thermal analysis); 
moreover, the meson width (enhanced {\it in medio}) softens the singularity and needs to be accounted for, so  $\left(\! q^2\!-\!m^2\!+\!i\epsilon \right)^{-1}\! \rightarrow\! \left[\!2 E\! \left(\!q^0\! -\! E\! +\!i \frac{\Gamma}{2} \right)\!\right]^{-1} $ with the on-shell energy 
$E =\sqrt{\vec{q}^2 + m ^2 }$. This yields, for kinematics near the singularity,
\begin{eqnarray}
I_{\triangleleft} \simeq    i \int \frac{d^4 q}{(2\pi)^4}\frac{1}{8 E_1 E_2 E_3 }  \label{loop2}
\frac{1}{\left((P^0-q^0) - E_1(q) + i \frac{\Gamma_1}{2} \right)}  \\ \nonumber 
\frac{1}{\left[q^0-E_2(q)+i \frac{\Gamma_2}{2} \right]
\left[(P^0-q^0-k^0)-E_3(q)+i \frac{\Gamma_3}{2} \right]}
\end{eqnarray} 
where $E_{i=1,2} \equiv E_i (\vec{q}) = \sqrt{\vec{q}^2 + m_i ^2 }$,  $ E_3 \equiv E_3(\vec{q} + \vec{k}) = \sqrt{\vec{q}^2 + \vec{k}^2 + 2 |\vec{q}||\vec{k}| z + m_3 ^2 }$. 
The complex energy (pole position)
$\tilde E_j := E_j-i\frac{\Gamma_j}{2}$ helps shorten notation.
The finite-temperature loop is calculated in the Matsubara formalism, in which the $q^0$ integral of Eq.~(\ref{loop2}) is replaced by a sum over Matsubara frequencies
\begin{eqnarray}
q^0 \rightarrow i \omega _{n }  =   i \frac{2\pi n }{\beta } &\ & \ \ n = 0,\pm 1 , \pm 2, \cdots, \label{Matsubara} \\
\int\frac{d q^0}{(2\pi)}
f(q^0,\vec{q})   \rightarrow S_M & = & \frac{i}{\beta }\sum_{ n = -\infty}^{\infty} f \left( \omega _{n }, \vec{q} \right), \label{feynmanrule}
\end{eqnarray}
and $\beta:=1/T$. The Matsubara sum can be analytically carried out: $S_M$ is first reinterpreted as a contour integration around the OY axis in the complex $x$ plane, $S_M\to \frac{1}{2\pi}\oint dx \frac{1}{2} \coth (\beta x/2)$. The contour is then deformed, picking up the poles of the integrand~\cite{Nieto:1993pr}, that substitute the arguments of the $\coth$.

Before doing this, $P^0$ and $k^0$ also have to be analytically continued to $iP^0$ and $ik^0$, ensuring that none of the poles fall along the imaginary $q^0$ axis upon cancelling, {\it e.g.} $P^0-E(q)$, which would spoil this Sommerfeld method yielding the hyperbolic cotangent. Once all the poles of the denominators have correctly been picked up, $P^0$ and $k^0$ can be continued back to Minkowski space~\cite{Aurenche:1991hi,Evans:1991ky,Laine:2016hma}.

The $\coth$ is given physical interpretation~\cite{Weldon:1983jn} in terms of boson emission to the thermal medium by the Bose-Einstein (BE) occupation function $n_\beta(E)$ in $\coth(\beta E/2) 
=[1+2n_\beta(E)]=(1+n_\beta)^2-n_\beta^2$.   
(A slight modification occurs for fermions, with $\tanh(\beta E/2)$ yielding the Fermi-Dirac combination $(1-n_\beta)$ instead.)
The thermal triangle loop becomes 
\begin{eqnarray}    
\label{loop-finiteT} 
I_{\triangleleft} & \simeq &   \frac{1}{2} \int \frac{d^3 q}{(2 \pi)^3} \frac{1}{8 E_1 E_2 E_3} 
\frac{1}{\left(P^0 - \tilde E_1 -\tilde E_2 \right)} \nonumber \\
& \times &   \frac{1}{\left( P^0 - k^0 - \tilde E_2 - \tilde E_3 \right)}   \frac{1}{\left( k^0 - \tilde E_1 + \tilde E_3 \right)}  \\   \nonumber
& \times & \left\{ \left[ 1 + 2n_\beta(\tilde E_2) \right] \left(\!  -k^0 \!+\!  \tilde E_1\! -\! \tilde E_3\! \right) \right. 
\nonumber \\ &\phantom{\{}  & +
\left[ 1+2n_\beta(P^0- \tilde E_1 ) \right] \left(\! P^0\! -\! k^0\! -\!  \tilde E_2\! -\!  \tilde E_3\! \right)  \nonumber \\ &\phantom{\{}  & + \left.
   \left[ 1+2n_\beta ( k^0 - P^0 + \tilde E_3) \right] 
\left(\!P^0\! -\! \tilde E_1\! -\! \tilde E_2\! \right) \right\}
\end{eqnarray}
%
where each term contains two usual propagators and the third one leaves a $(1+2n_\beta)$ factor ``lasing''  that third particle into the medium (to obtain the vacuum result, it suffices to set $n_\beta=0$). For heavy mesons, whose density is small at $T\sim $ sub-GeV, this is not the main reason for the singularity waning.
The small $\Gamma_i$ (and thus the $E_i\leftrightarrow \tilde E_i$ distinction) can be ignored in the 
BE factors (analytical for physical masses) but need to be tracked down in the denominators of the propagators.

\newpage 

\section{Numerical computations}
\label{Results}

\subsection{A singularity appearing in $\Lambda (1405)$ production} \label{subsec:lambda}

\begin{figure}[h]
\centering
\includegraphics[width=0.8\columnwidth]{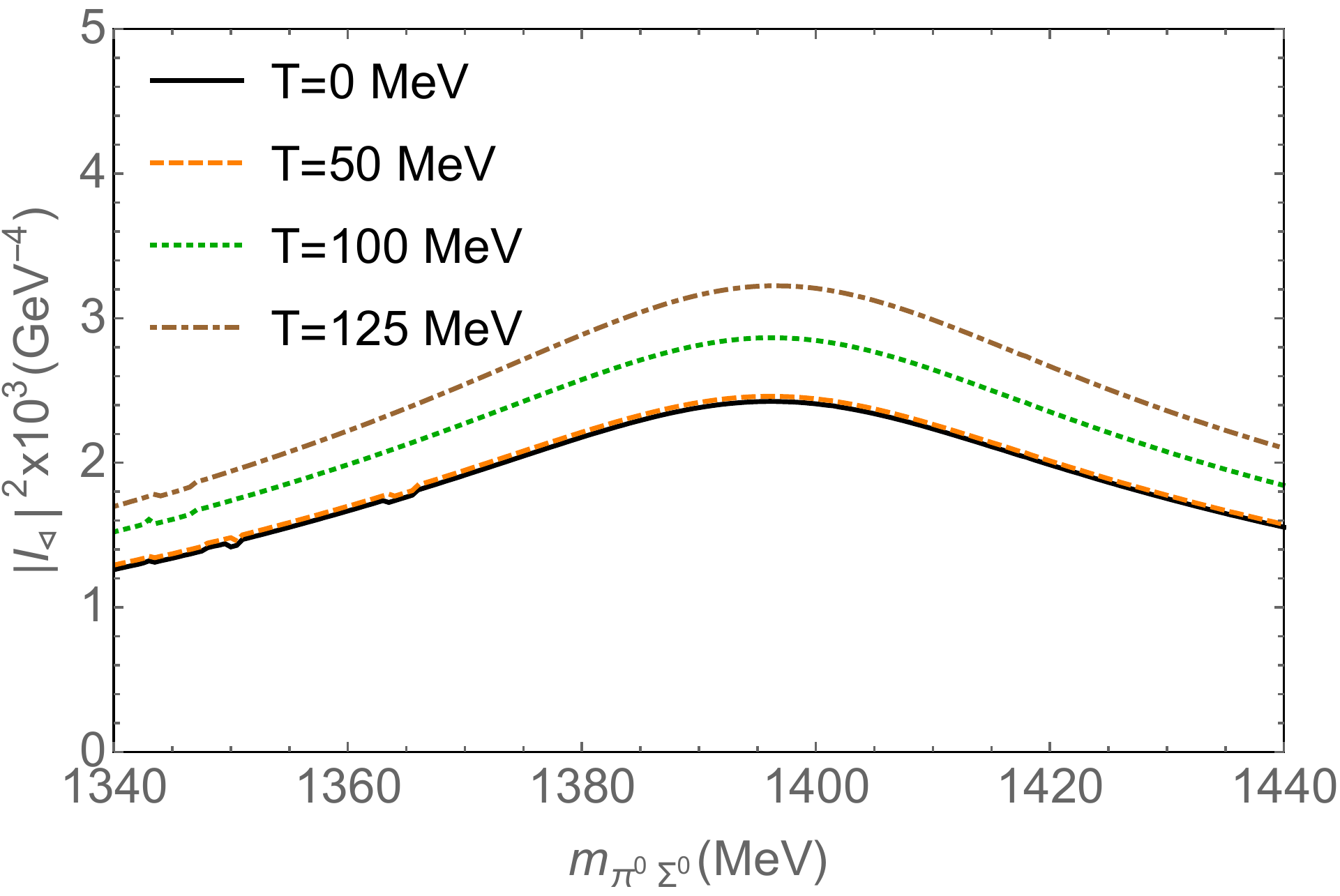} 
\caption{
\label{fig:triangle-KSTARSIGMAPION}
Squared modulus of triangle loop integral $(|I_\triangleleft|^2)$ in Eq.~(\ref{loop-finiteT}) for the reaction~(\ref{reaccao2}) as function of $m_{\pi^0 \Sigma^0}$~\cite{Bayar:2017svj} at finite $T$, with
intermediate particles 1, 2 and 3 being  $K^{\ast 0}, \Sigma^{ 0}, \pi^{0} $ respectively. The invariant mass of  $K^{* 0} \Sigma^{ 0} $ is fixed at 2140 MeV. Masses and widths taken at their vacuum values~\cite{Tanabashi:2018oca} except for a regulating $\Gamma _2, \Gamma _3 = 0.2$ MeV to avoid numerical instability (so the peak, $\propto \log \Gamma$ \cite{Debastiani:2018xoi}, is underestimated).}
\end{figure}

As a first application, we examine the triangle singularity coming from the $K^{\ast} \Sigma \pi $ triangle diagram, important for $\Lambda (1405)$ production in $\pi^- p \rightarrow K^{0} \pi \Sigma $ and $p p \rightarrow p K^{+} \pi \Sigma $, discussed in detail in~\cite{Bayar:2017svj}.
The triangle diagram arises from the formation of an $N^\ast$ resonance decaying, in the notation of fig.~\ref{fig:triangle}, by
\begin{eqnarray}\label{reaccao2}
N^\ast ({\rm A})  \rightarrow \underbrace{K^{\ast} (1)}_{\to K({\rm B}) \pi(3)}& & \Sigma (2) \\ \nonumber
 & & \xrightarrow{(2)+(3)}   \Lambda(1405) ({\rm C}) .
\end{eqnarray}
Unlike the next couple of examples to follow, a pion is present in the triangle diagram: because 
$m_{\pi} \sim T$ is reachable the virtual pion can be taken/deposited on shell in the medium with little or no Boltzmann suppression. 
To isolate this enhancing effect of the $(1+2n_{\beta})$ factor in Eq.~(\ref{loop-finiteT}) 
we fix masses and widths to their vacuum values and show
$|I_\triangleleft|^2$ against the near-threshold $\pi^0 \Sigma^0 $ invariant mass 
in Fig.~\ref{fig:triangle-KSTARSIGMAPION}.

This mild singularity is not strongly affected by the medium because the intermediate particles are not very affected themselves. The next example will clarify this further.

\subsection{A singularity relevant for $a_1(1420)$ production} \label{subsec:a1}

\begin{figure}
\centering
\includegraphics[width=0.8\columnwidth]{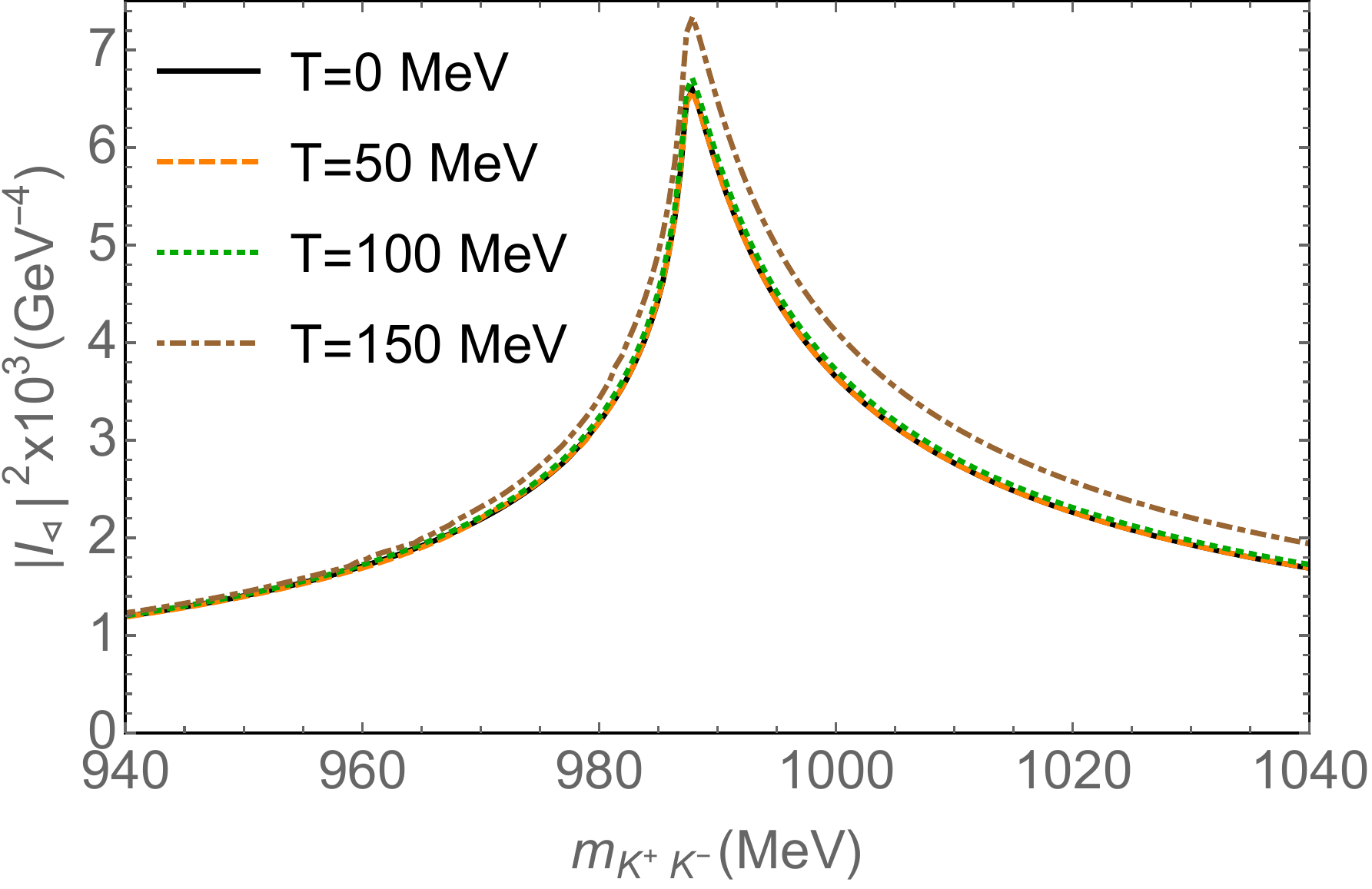} \\
\includegraphics[width=0.8\columnwidth]{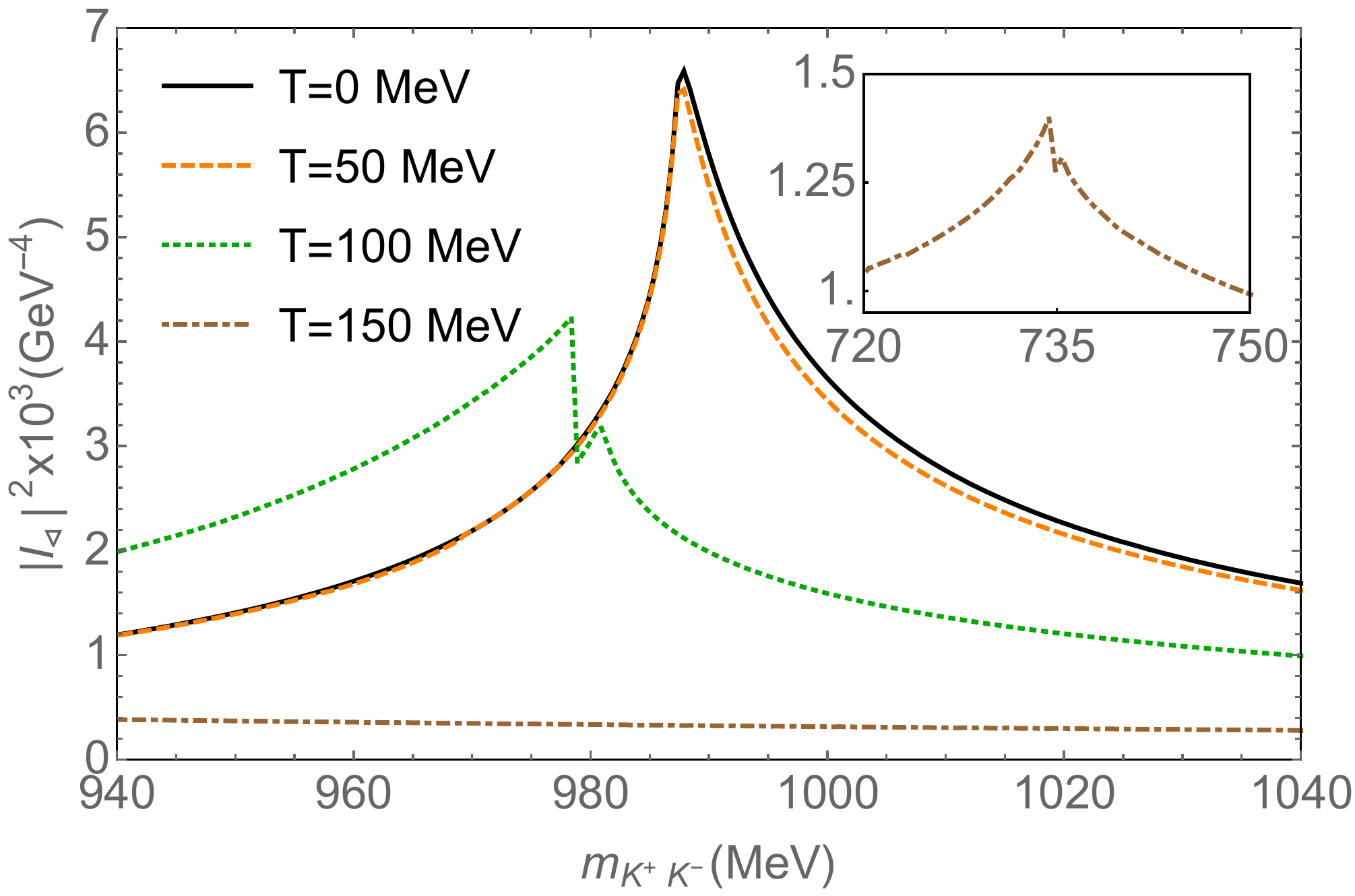} 
\caption{Squared scalar triangle loop integral $(|I_\triangleleft|^2)$ in Eq.~(\ref{loop-finiteT}) for the reaction of Eq.~(\ref{reaccion2}), as a function of the $ K^+ K^-$ invariant mass~\cite{Aceti:2015zva,Guo:2019twa}, for various temperatures. The intermediate particles 1, 2 and 3 are respectively $K^{*+},  K^{-}$ and $ K^+$,  with fixed $K^{*+} K^- $ invariant mass at 1420 MeV.
{\bf Top}: temperature included only in the loop variable $q_0$, but $M_i$, $\Gamma_i$  fixed at their vacuum values~\cite{Tanabashi:2018oca} except for a regulating $\Gamma _2 = 0.5$ MeV to avoid numerical instability (so the $T=0$ peak is estimated from below). The effect of $T$ is smaller than in figure~\ref{fig:triangle-KSTARSIGMAPION}, due to $m_K>>m_\pi\sim T$. 
{\bf Bottom}: thermal corrections to the $(1,2,3)$ meson masses and widths are included, according to the values shown in Table~\ref{thermal-masses-widths}. The singularity melts with $T$. 
}
\label{fig:triangle-KSTARKK}
\end{figure}

As a second example we consider the singularities in the pionless $K^{*}  \overline{K} K$ triangle diagram,  relevant in discussing  resonances like the $a_1(1420)$ reported by COMPASS in the $P-$wave $\pi f_0(980)$ channel of the $\pi p \rightarrow \pi \pi \pi p$ reaction~\cite{Guo:2019twa,Adolph:2015pws}. To study the $K^{\ast} K$ threshold area, we analyze simplified processes such as 
\begin{eqnarray} \label{reaccion2}
{\rm A}\! \rightarrow\! K^{*+}(1)  K^{-}(2)  K^+ (3)\! \rightarrow \!
\pi^0 ({\rm B}) 
(\pi^+ \pi^-) / (\pi^0 \eta)
({\rm C}) .
\end{eqnarray}

\noindent
As seen in the top panel of Fig.~\ref{fig:triangle-KSTARKK}, unlike the previous example with a pion, the inclusion of thermal effects through the Matsubara prescription in the triangle loop integral alone does not induce a relevant change because $m_K>T$ (this is certain to change at the yet higher temperatures in the quark-gluon plasma where the $\coth$-function is modified, with $T$ more comparable to the masses). 
In contrast, when thermal corrections to the masses and widths are included according to Table~\ref{thermal-masses-widths}, the triangle singularity is strongly affected (bottom plot). The narrow peak, appearing at smaller temperatures, becomes smaller and moves to the left due to the reduction of $K^{+} K^- $ threshold.  
It also splits into two, probably corresponding to the separation of the threshold and triangle singularities as explored in the next example.
This is a curious feature of the line shape in the bottom panel. In vacuum, the Landau singularity usually appears at the threshold or above it. Maybe this is due to the thermal medium (that can provide energy and move a phenomenon below nominal threshold) or simply a numerical effect, as we find some sensitivity to the parametrized masses $M(T)$ and widths $\Gamma(T)$. In any case, it does not affect our statement that the intensity of the loop function is very depressed upon increasing the temperature.

\begin{center}
\begin{table}[h]
\caption{Input meson thermal masses and widths~\cite{Cleven:2017fun,Gu:2018swy,Montana:2020lfi} for increasing temperatures, all in GeV.
We exclude $m_\pi$, practically constant and with modification of unclear sign~\cite{Schenk:1993ru}.
We are not currently aware of a detailed computation of the $K^*$ width, so we use a crude estimate of order $\Gamma_{K^*}(T=0)\times \left(1+ n_{\pi}(T) + n_{K}(T)\right)$.
} \begin{center}
\vskip1.5mm
\label{thermal-masses-widths}
\begin{tabular}{c|c|c|c|c}
\hline
\hline
T & $ 0$ & $ 0.05$ & $ 0.1$ & $ 0.15$   \\ 
\hline
$m_{K^{\pm}} $ & 0.49367 & 0.49367 & 0.4906 & 0.37  \\ 
$m_{K^{\ast +}} $ & 0.89166 & 0.8877 & 0.8207 & 0.508 \\
$m_{D^{\ast -}} $ & 2.0103 & 2.0099 & 1.994 & 1.872 \\
$m_{D^{\ast 0}} $ & 2.00685 & 2.00647 & 1.991 & 1.868  \\
$m_{D^{0}} $      & 1.86483 & 1.86466 & 1.856 & 1.776 
\\
$\Gamma_{K^{\ast +}} $ & 0.0508 & 0.0509 & 0.0532 & 0.0588  \\
$\Gamma_{D^{\ast -}} $ & $83.4 \times 10^{-6}$ & $87.6 \times 10^{-6}$
 & 0.00787 & 0.0359 
\\
$\Gamma_{D^{\ast 0}} $ & $55 \times 10^{-6}$ & $ 57.8 \times 10^{-6}$ & 0.0052 & 0.0237 
\\
\hline
\hline
\end{tabular}
\end{center}
\end{table}
\end{center}

\subsection{A debatable singularity in specific configurations of $X$ production} \label{subsec:X1}

Our next example concerns a singularity recently reported~\cite{Nakamura:2019nwd} {\it in lieu} of the $X(3872)$ in the reaction 
\begin{eqnarray}\label{reaccao1a}
B^0  \rightarrow K^+ + &\overbrace{\left( D^{*-}(1) D^{*0}(2) \right)}^{({\rm A})}&
\\ \nonumber
 & \xrightarrow{D^0 (3)}  &\!\!\!\!\!\!\!  K^+ + \overbrace{\pi^-}^{({\rm B})}
\overbrace{\left( J/\psi \pi^+ \pi^-  \right)}^{({\rm C})} .
\end{eqnarray}
In Heavy Ion Collisions, because the expansion time is of order $10\ {\rm fm}/c$ and the $B$ meson decays weakly outside of the medium, this triangle could not be initiated by a $B\to K + \dots$ decay. Instead, the relevant question is whether \emph{prompt} $X$ production
\footnote{The CMS collaboration was able to use its vertex tracker to potentially separate its  $X$ sample into prompt (from direct $c\bar{c}$ production) and non prompt (from delayed weak decays of the $B$ meson) subsets~\cite{CMS:2019vma}.}
of the $X$ meson (sampling those $X(3872)$ produced within the fireball from $c\bar{c}$ pairs) can be contaminated by the triangle singularity,
\begin{eqnarray}\label{reaccao1}
c\bar{c}  \rightarrow &\overbrace{\left( D^{*-}(1) D^{*0}(2) \right)}^{({\rm A})}&
\\ \nonumber
 & \xrightarrow{D^0 (3)}  &\!\!\!\!\!\!\!   \overbrace{\pi^-}^{({\rm B})}
\overbrace{\left( J/\psi \pi^+ \pi^-  \right)}^{({\rm C})} .
\end{eqnarray}
Our findings, developed next, show that, even if the precise kinematic conditions are met (and in medium production this will be the case, as all available phase space is sampled with some probability), the singularity is erased in an infinite medium.

The integral $|I_\triangleleft|^2$ is calculated as function of the invariant mass of $J/\psi \pi^+ \pi^-$, 
taking the mass of the final system, coincident with that of the initial $c\bar{c}$ pair, fixed at 4.0172 GeV, which is the value that would trigger the accidental singularity in vacuum.

\begin{figure}[h]
\centering
\includegraphics[width=0.8\columnwidth]{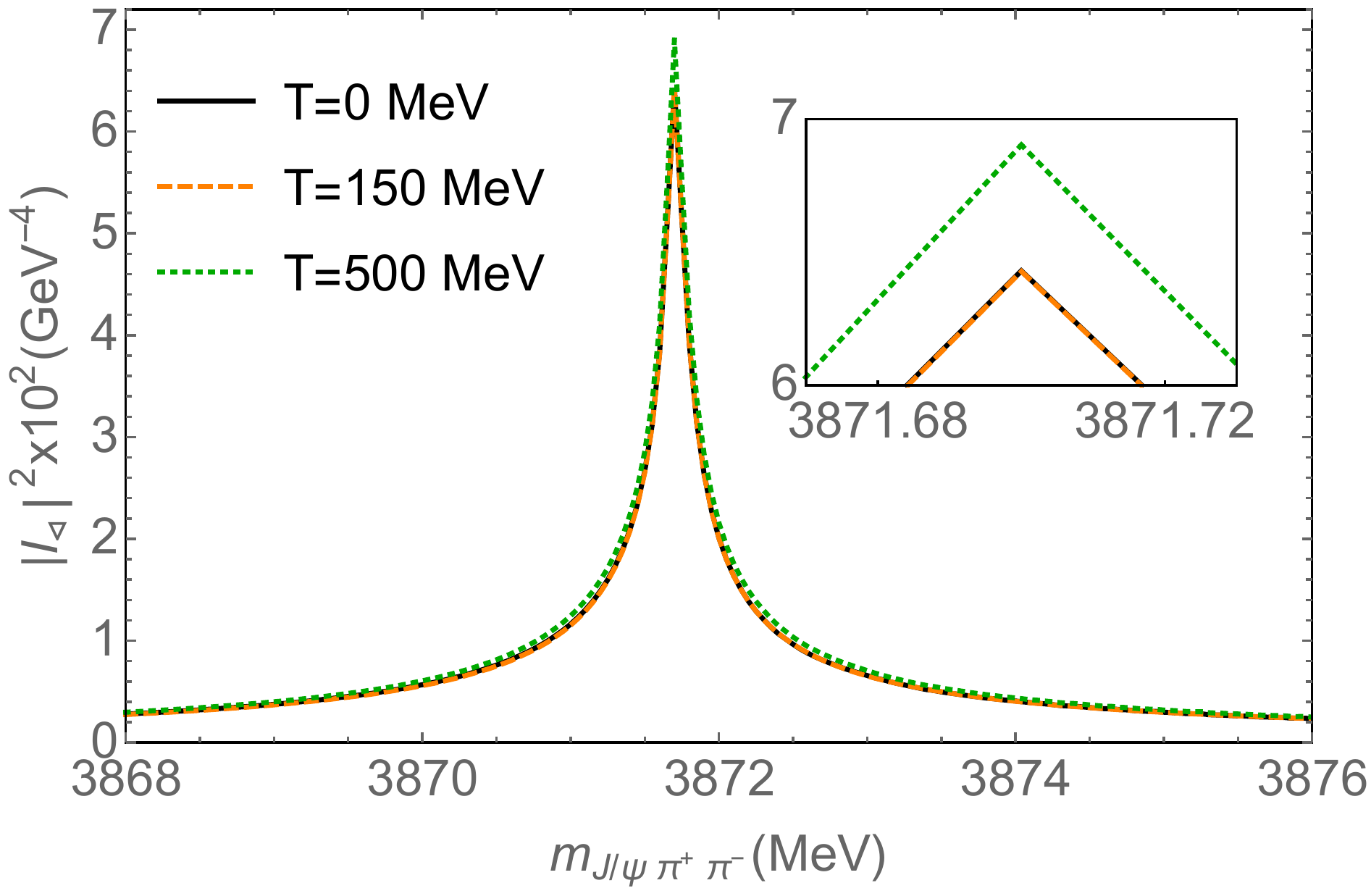} \\
\includegraphics[width=0.8\columnwidth]{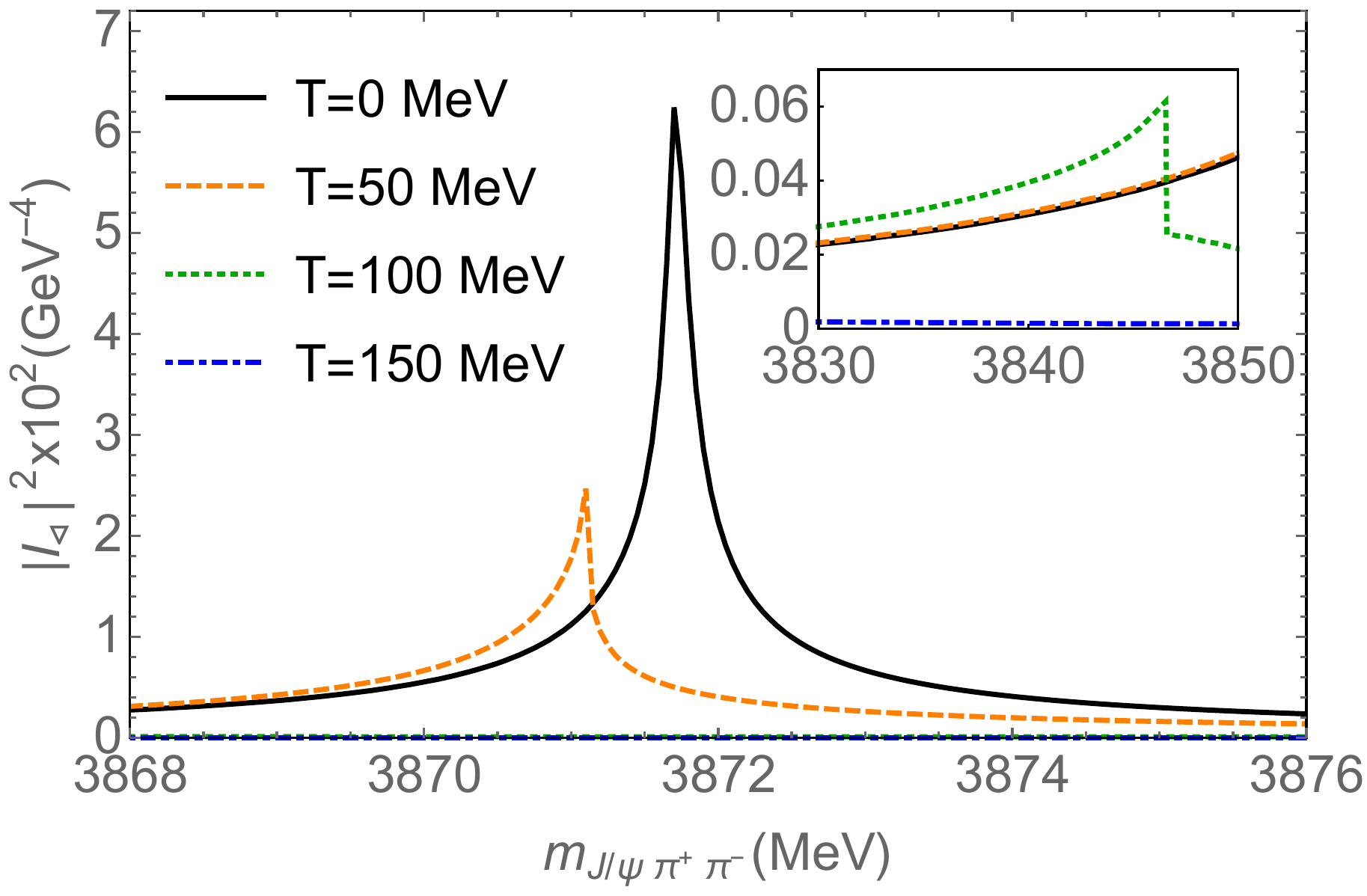} 
\caption{Squared triangle graph $(|I_\triangleleft|^2)$ in Eq.~(\ref{loop-finiteT}) against $m_C$ for $c\bar{c}(A) \rightarrow  (J/\psi \pi^+ \pi^-)(C) \pi^-(B)$~\cite{Nakamura:2019nwd}.
{\bf Top}: the temperature affects only the loop variable, but $M_i$, $\Gamma_i$ are fixed at their vacuum values~\cite{Tanabashi:2018oca}. The effect of $T$ is very small.
{\bf Bottom}: thermal corrections to the $(1,2,3)$ meson masses and widths included  (Table~\ref{thermal-masses-widths}); the singularity melts with $T$.
}
\label{fig:triangle-DSTARDSTARD}
\end{figure}

This triangle with heavy quarks is even less affected itself 
than the case of $K^{*}  \overline{K} K$ 
when the thermal effects are implemented through Matsubara's prescription alone, due to the insufficient temperatures, as we plot it in figure~\ref{fig:triangle-DSTARDSTARD} (top panel). 
We next assess the temperature effect on each individual $D$ meson
 participating in the triangle loop. The leading thermal effect is an increase of the resonance width~\cite{Weldon:1993dj,Dobado:2002xf} due to Bose-Einstein enhancement of the bosonic channels to which it can decay, but with increasing temperature the particle masses are also affected.
This $m_i(T)$, $\Gamma_i(T)$ behavior is already known~\cite{Cleven:2017fun,Gu:2018swy,Montana:2020lfi}
for the $\pi, D^{*-},  D^{*0}, D^0$ needed in our examples (table~\ref{thermal-masses-widths}; for  $\Gamma_{K^*}$ we provide an educated guess), 
and we can directly study its influence on the triangle $I_\triangleleft$.

We plot the resulting $|I_\triangleleft|^2$ in Fig.~\ref{fig:triangle-DSTARDSTARD}. 
As the $D/D^*$ mesons decay to lighter particles $(\pi,K)$ with large in-medio occupation number, the singularity is molten away. In fact, we do find very acute sensitivity to the kinematic variables. 
The peak of the singularity is sharpest at $T = 0$:  as temperature increases, the peak diminishes in height and shifts to smaller $ J/\psi \pi^+ \pi^-$ invariant mass $m_C$, accompanying the drop of the $D^{*0} D^0$ threshold. At even higher $T$, the structure seems to disappear altogether (and is not made visible even changing $m_A$).

\begin{figure}[h]
\centering
\includegraphics[width=0.8\columnwidth]{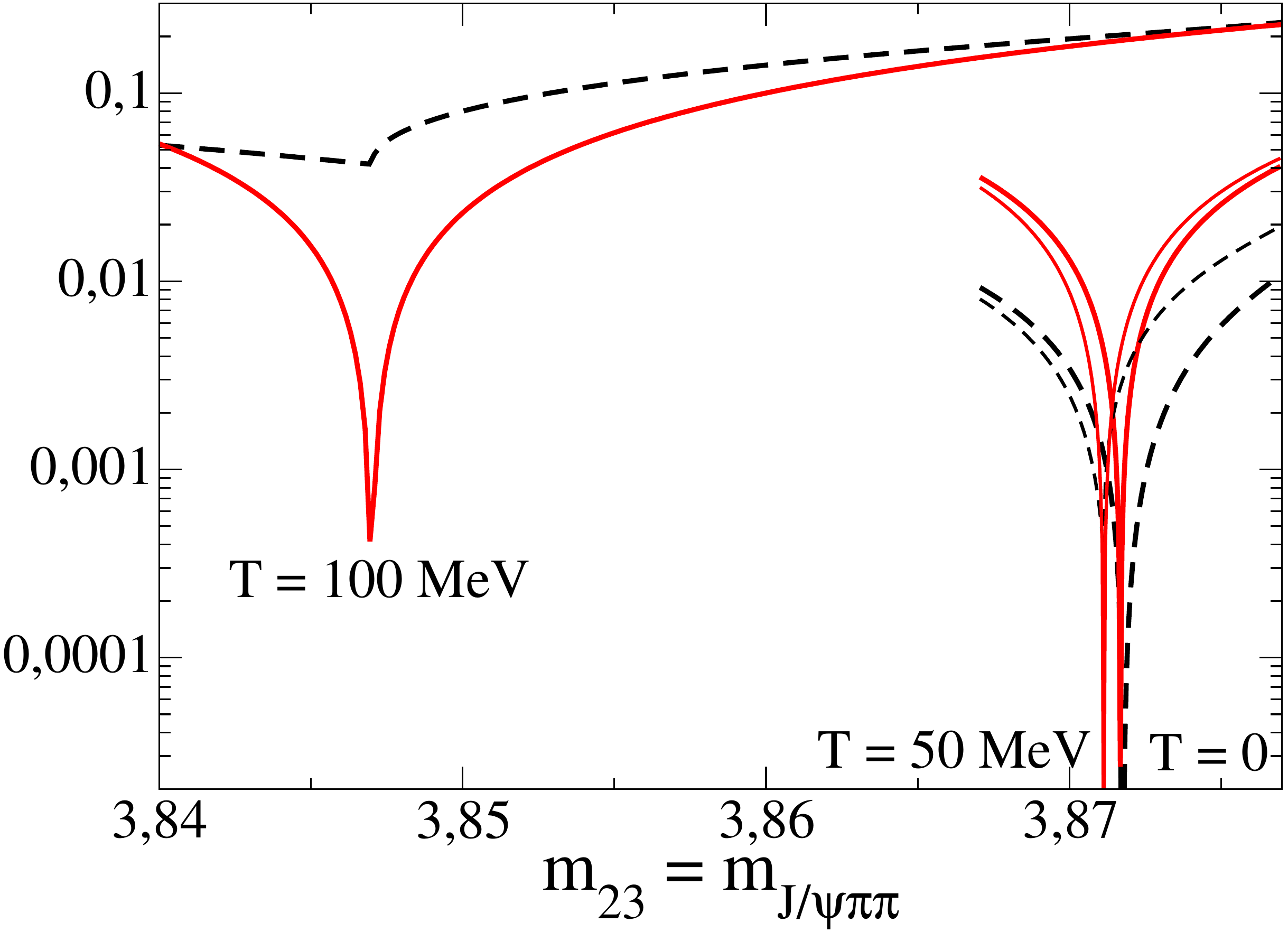} 
\caption{Kinematic singularity conditions for the triangle (red solid line) $\lim_{\epsilon \rightarrow 0}{(q_{on +} - q_{a - })} = 0$, and two-body threshold  (black dashed line), $\lim_{\epsilon \rightarrow 0}{(q_{a + } - q_{a - })}$, as functions of $m_{23} = m_{J/\psi \pi \pi}$ invariant mass, at different temperatures.
\label{fig:kin-cond}} 
\end{figure}

To better understand this disappearance, we examine the reaction kinematics following~\cite{Guo:2019twa,Bayar:2016ftu}. The position of the triangle singularity in the invariant mass $m_{23} \equiv m_C$, is determined by finding a solution of the pole-pinch condition putting the three intermediate particles on-shell and in the collinear kinematics that allows them to interact classically (with particle 3 reaching particle 2),
\begin{eqnarray}\label{eq:kin_cond2}
& & \lim_{\epsilon \rightarrow 0}{(q_{\mathtt{on} +} - q_{a - })} = 0, \\  \nonumber 
q_{\mathtt{on} \pm} & := & \sqrt{\lambda(m_A ^2 , m_1 ^2 , m_2 ^2)} \pm i \epsilon , \\  \nonumber 
q_{a \pm} & := & \frac{[ k (m_{23} ^2 + m_2 ^2 - m_3 ^2) \pm \frac{\sqrt{\lambda(m_{23} ^2 , m_2 ^2 , m_3 ^2)}}{E_{23}}]}{2m_{23}^2} \pm i\epsilon;
\end{eqnarray}  
with $E_{23} \equiv E_C$. 
Fig.~\ref{fig:kin-cond} then shows the satisfaction of the kinematic Eq.(~\ref{eq:kin_cond2})
for the triangle $D^{*}  D^{*} D$ as function of $m_{23}$. 
Also plotted is the satisfaction of the condition for the two-body threshold singularity, 
\begin{equation} 
\lim_{\epsilon \rightarrow 0}{(q_{a + } - q_{a - })}=0\ .
\end{equation}
At $T=0$ both conditions are simultaneously satisfied (the lines touch the $x$-axis together), 
so the integration path in Eq.~(\ref{loop2}) is pinched between $q_{\mathtt{on} +}\to q_{a +}$ 
and $q_{a-}$. As $T$ increases, they shift to smaller $m_C$
 separating among themselves and from the $x$-axis, so the singularities wane in agreement with Fig.~\ref{fig:triangle-DSTARDSTARD}.

\subsection{A triangle singularity in $X(3872)\to \pi\pi\pi$
} \label{subsec:X2}

Now we examine another triangle singularity that appears in the decay
 $X(3872)\to \pi^0\pi^+\pi^-$ due to the intermediate~\cite{Molina:2020kyu} $D^* DD$ triangle. 
The interest in this singularity is the possibility to use it to precisely determine the mass 
of the $X\sim \chi_{c_1}(3872)$ meson, following earlier ideas~\cite{Guo:2019qcn,Braaten:2019gfj}
on radiative decays of the $X$ that would be less germane to heavy ion collisions. 
The difference with the example in subsection~\ref{subsec:X1} 
 is that the triangle singularity 
appears in the \emph{decay} and not in the \emph{production} of the pseudovector $X$ meson.

\begin{figure}[h]
\includegraphics[width=0.8\columnwidth]{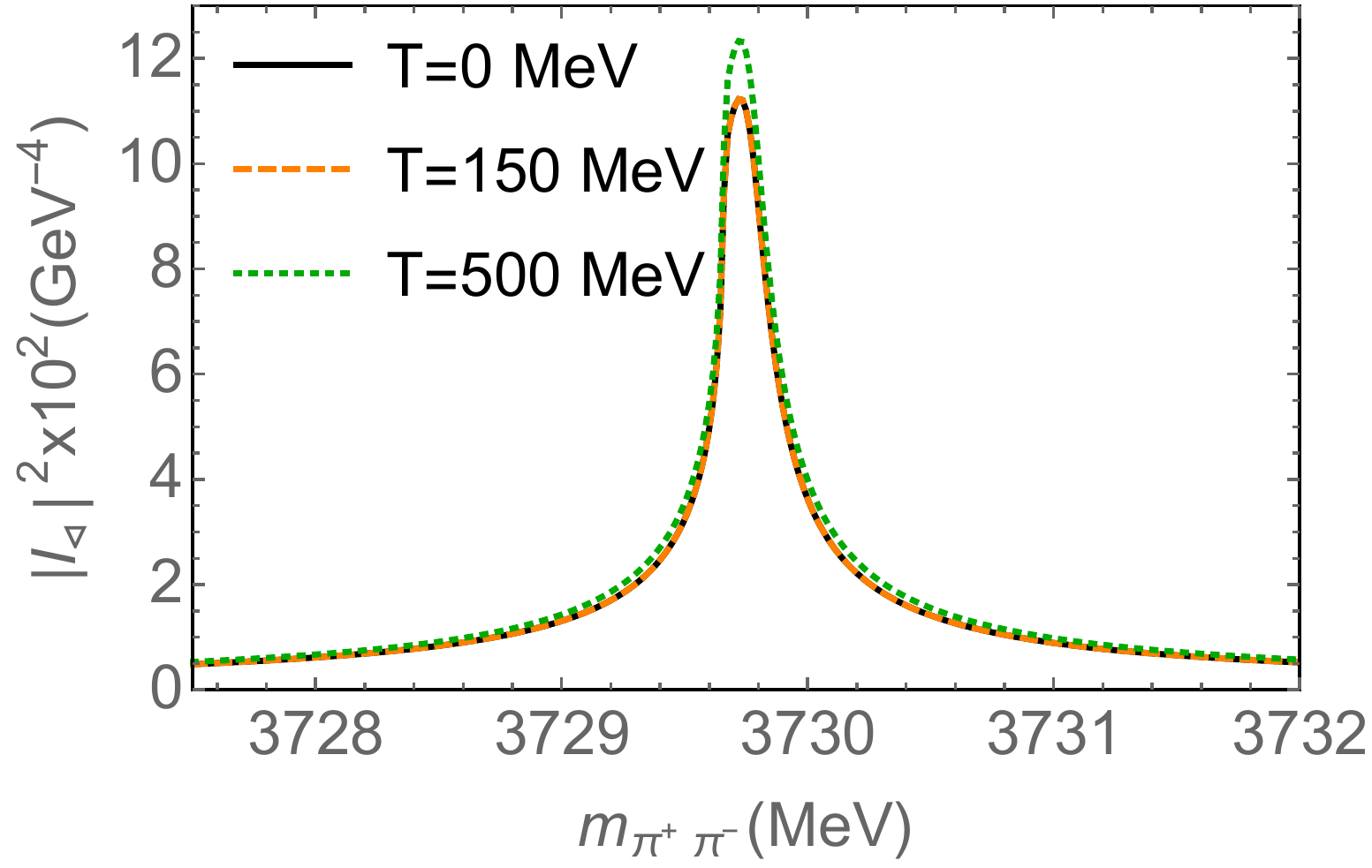} \\
\includegraphics[width=0.8\columnwidth]{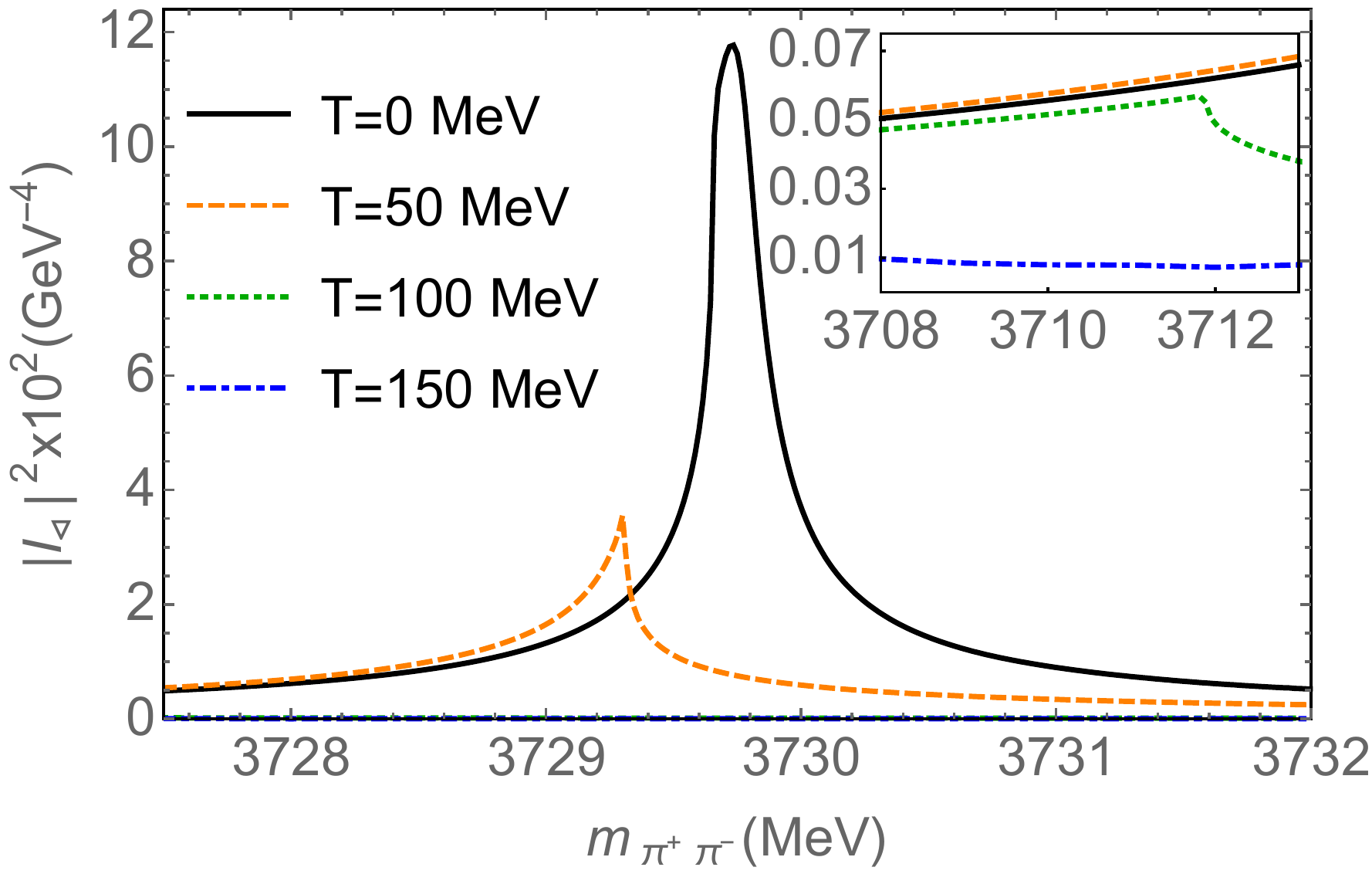}
\caption{
The triangle diagram that would appear in the $\pi\pi$ spectrum from $X\to 3\pi$ decays is not very much affected by temperature (top), even at a large 500 MeV temperature that of course puts the computation outside the realm of hadrons and into the qgp, requiring other treatment (but it is shown for illustration). However, if thermally modified masses and widths are included, the singularity is quickly washed away. 
\label{fig:Xto3pi}
}
\end{figure}

Figure~\ref{fig:Xto3pi} shows how the singularity is very much affected by the thermal width and modified thermal mass of the particles involved (bottom panel), whereas the triangle diagram itself, not so much (top panel).

Because the fireball has a finite lifetime, the most interesting examples for initial studies would be those cases with broader widths, that could immediately be ruled out to be triangle singularities. The narrower ones, for example this one involving the $X$ meson, will require more detailed simulation since part of the  $X$ decays will occur already after the medium has undergone kinetic freeze out (that is, local equilibrium can no longer be maintained by the pion interaction rate).  
\newpage
\newpage
\subsection{A singularity appearing in deuteron  production}  \label{subsec:dstar}

A last example is related to a reaction involving an excited state of the deuterium nucleus, $d^*(2380)$. Because the $pn$ deuteron bound state is the simplest nucleus, it is a benchmark for nuclear interactions, and the existence (or not) of a resonance thereof~\cite{Adlarson:2014pxj} is of great importance to nuclear physics. 

A recent proposal explaining the J\"ulich detection but absence of evidence from other studied reactions~\cite{Molina:2021bwp,Ikeno:2021frl} is that a triangle singularity arises because of a baryon loop with the particles $\Delta^+ p n $, that contributes to the fusion reaction $ p p  \rightarrow  \pi d  $.

This proceeds sequentially as $ p p  \rightarrow  \Delta (1232) N  $ followed by $ \Delta (1232)  \rightarrow  \pi N^{\prime}$, with the net result of a successful fusion of $ N N^{\prime} $ to form the deuteron; schematically we have 
\begin{eqnarray}\label{reaccao5}
p p ({\rm A})  \rightarrow \underbrace{\Delta (1232)^+ (1)}_{\to \pi^+({\rm B}) n(3)}& & p (2)
 \xrightarrow{(2)+(3)} \pi^+(B) d  ({\rm C}) .
\end{eqnarray}

\begin{figure}[h]
\centering
\includegraphics[width=0.8\columnwidth]{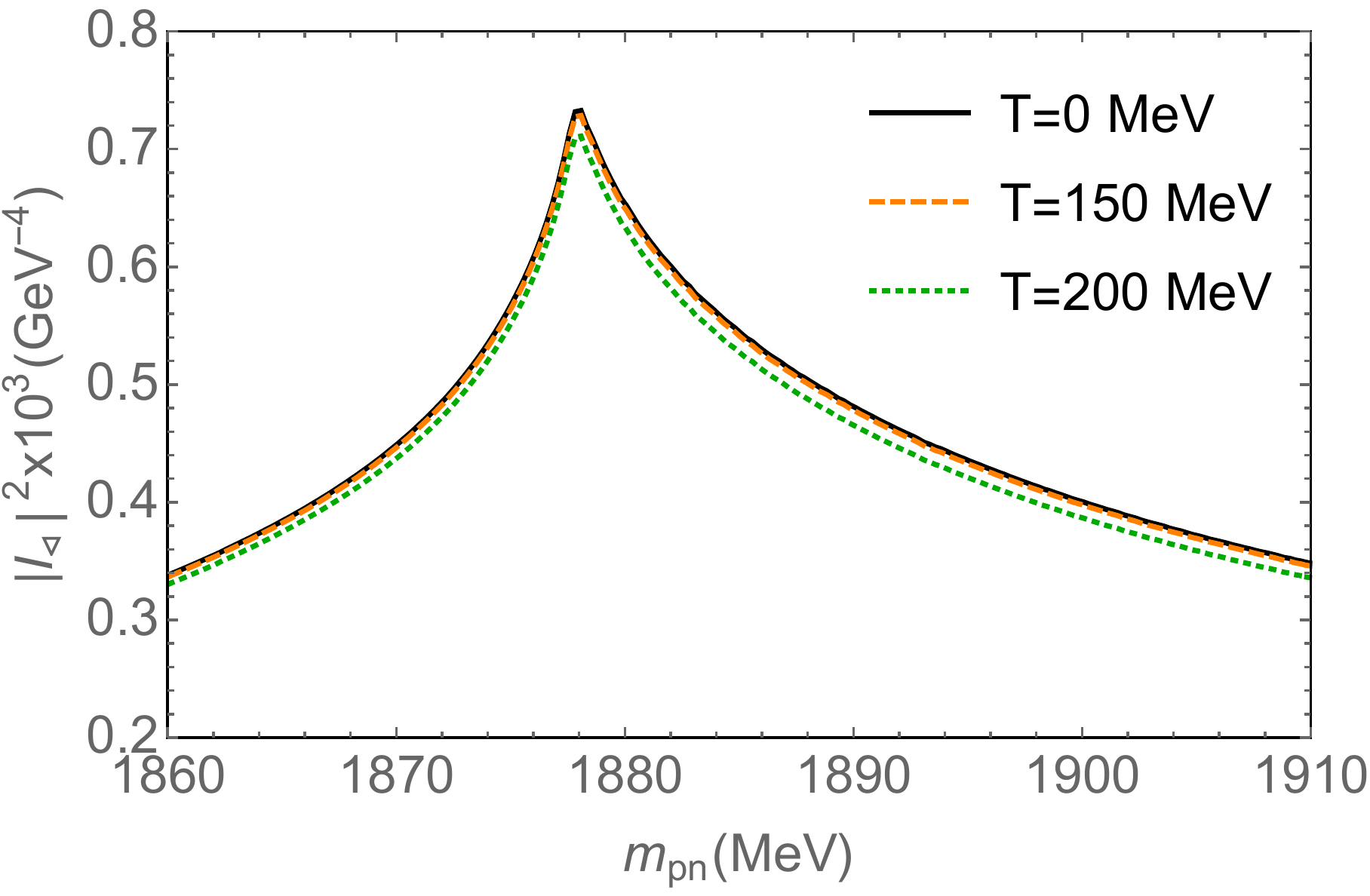} 
\\
\includegraphics[width=0.8\columnwidth]{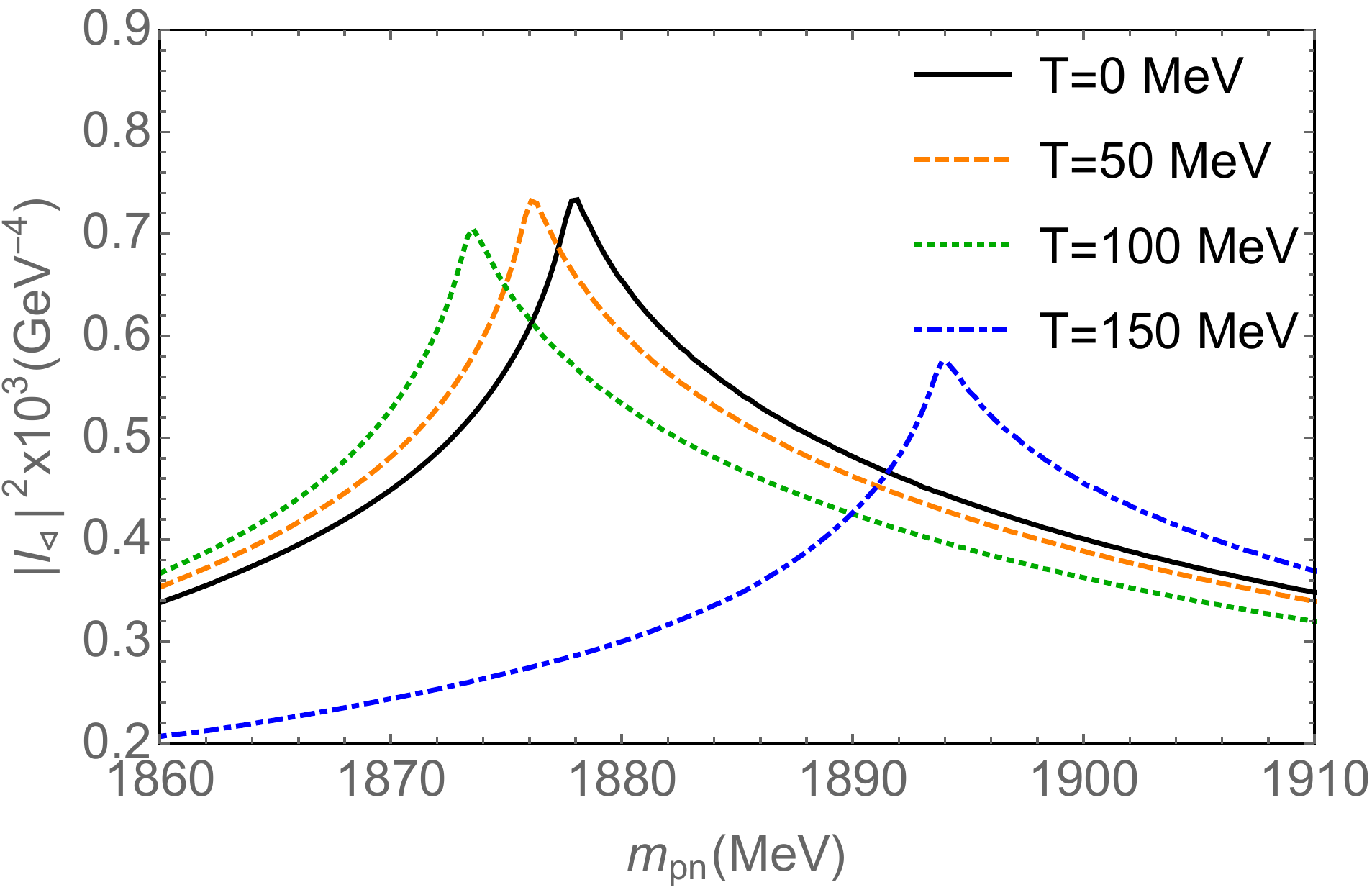} 
\caption{
\label{fig:triangle-DELTAPN}
Squared modulus of triangle loop integral $(|I_\triangleleft|^2)$ in Eq.~(\ref{loop-finiteT}) for the reaction~(\ref{reaccao5}) as function of $m_{p n}$ at finite $T$, with intermediate particles 1, 2 and 3 being  $\Delta^+ , p , n  $ respectively. The invariant mass of  $ \Delta^+ p $ is fixed at 2188.68 MeV. 
{\bf Top}: masses and widths taken at their vacuum values~\cite{Tanabashi:2018oca} except for a regulating $\Gamma _2 = 0.5$ MeV to avoid numerical instability.
{\bf Bottom}: the (small) thermal corrections to the $(1,2,3)$ meson masses and widths are included, according to the values shown in Table~\ref{thermal-masses-widths-baryons}. 
}
\end{figure}

Figure~\ref{fig:triangle-DELTAPN} displays the squared modulus of the triangle loop integral as a function of the $ p n $ invariant mass at different temperatures
(with the proviso that BE factors were substituted by Fermi-Dirac ones appropriate for spin-$1/2$ and $3/2$ particles).

As in earlier examples, the top plot has the masses and widths fixed to their vacuum values. In agreement with those examples including heavy particles, the inclusion of thermal effects through the Matsubara prescription in the triangle loop integral alone does not engender any sizeable change. 
In the bottom plot we attempt to include the thermal corrections to the masses for the particles involved in this reaction.  
We are aware of possible controversy on the choice of such thermal corrections; 
different results have been reported in the literature~\cite{Leutwyler:1990uq,Kacir:1995gy,AbuShady:2012zza,Azizi:2015ona,Azizi:2016ddw,Shang:2020kfc}. We have adopted the values in Table~\ref{thermal-masses-widths-baryons} but there should be no difficulty in changing them as needed.

\begin{center}
 \begin{table}[h]
 \caption{Input  $ p , n$ baryon thermal masses  (in GeV) extracted from~\cite{Kacir:1995gy}; input $\Delta^+ $ thermal width (in GeV) extracted from Eq. (21) of~\cite{Ghosh:2010md}, taking $g_{\Delta} = 2.38, \Lambda = 400 $ MeV and  the $T$-dependent values of $f_{\pi} $ according to~\cite{Kodama:1995kj}.  \label{thermal-masses-widths-baryons} }
 \begin{center}
 \vskip1.5mm
 \begin{tabular}{c|c|c|c|c}
\hline
\hline
T & $ 0$ & $ 0.05$ & $ 0.1$ & $ 0.15$   \\ 
\hline
$m_{p} $ & 0.9383 & 0.9373 & 0.9361 & 0.946  \\ 
$m_{n} $ & 0.9396 & 0.9387 & 0.9374 & 0.948 \\
$m_{\Delta} $ & 1.2349 & 1.2349 & 1.2349 & 1.2349 \\
$\Gamma_{\Delta} $ & 0.1311 &  0.1314 &  0.136 &  0.159 
\\
\hline
\hline
 \end{tabular}
 \end{center}
 \end{table}
\end{center}

Interestingly, the baryon masses and $\Delta(1232)$ width are affected very little by the bath at small $T$. This comes about because the thermal baryon population is Boltzmann suppressed, and the $\Delta$ is already so broad in vacuum that its width by pion decay does not substantially increase until quite high temperatures, at 150 MeV and beyond, nearing
what we think is the limit where one can discuss a hadron gas~\footnote{ A different result would be obtained if, additionally to finite T, we had considered a finite nucleon-number density (as appropriate for lower-energy collisions where some nucleons from the initial state do not fly away along the light cone but remain in the medium). Then, at a minimum, the $\Delta$ resonance would become broader and its mass would shift earlier~\cite{Azizi:2016ddw}. The triangle singularity in the corresponding finite-density evaluation would almost surely be erased.}. 

\begin{figure}[h]
\centering
\includegraphics[width=0.8\columnwidth]{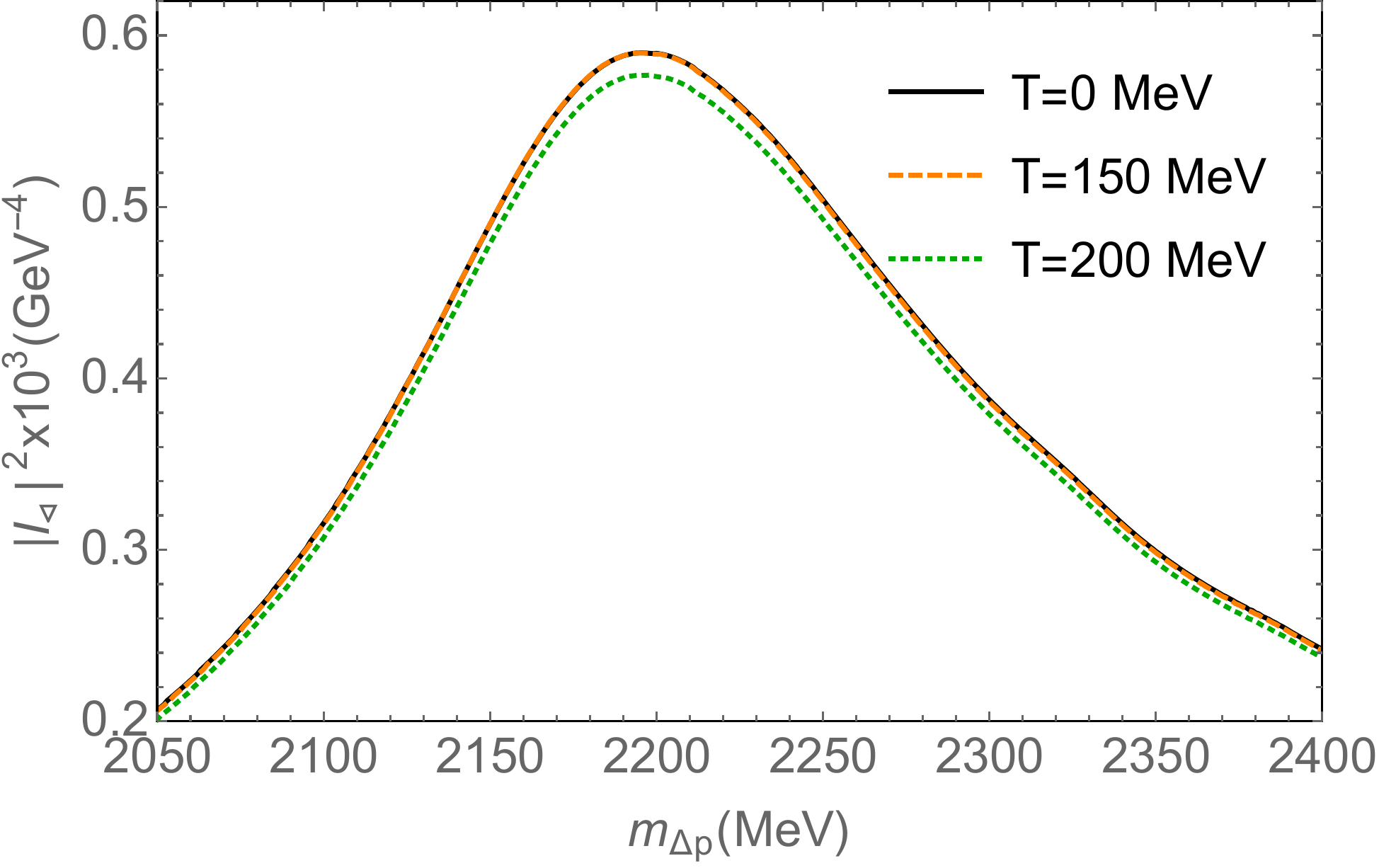}
\\
\includegraphics[width=0.8\columnwidth]{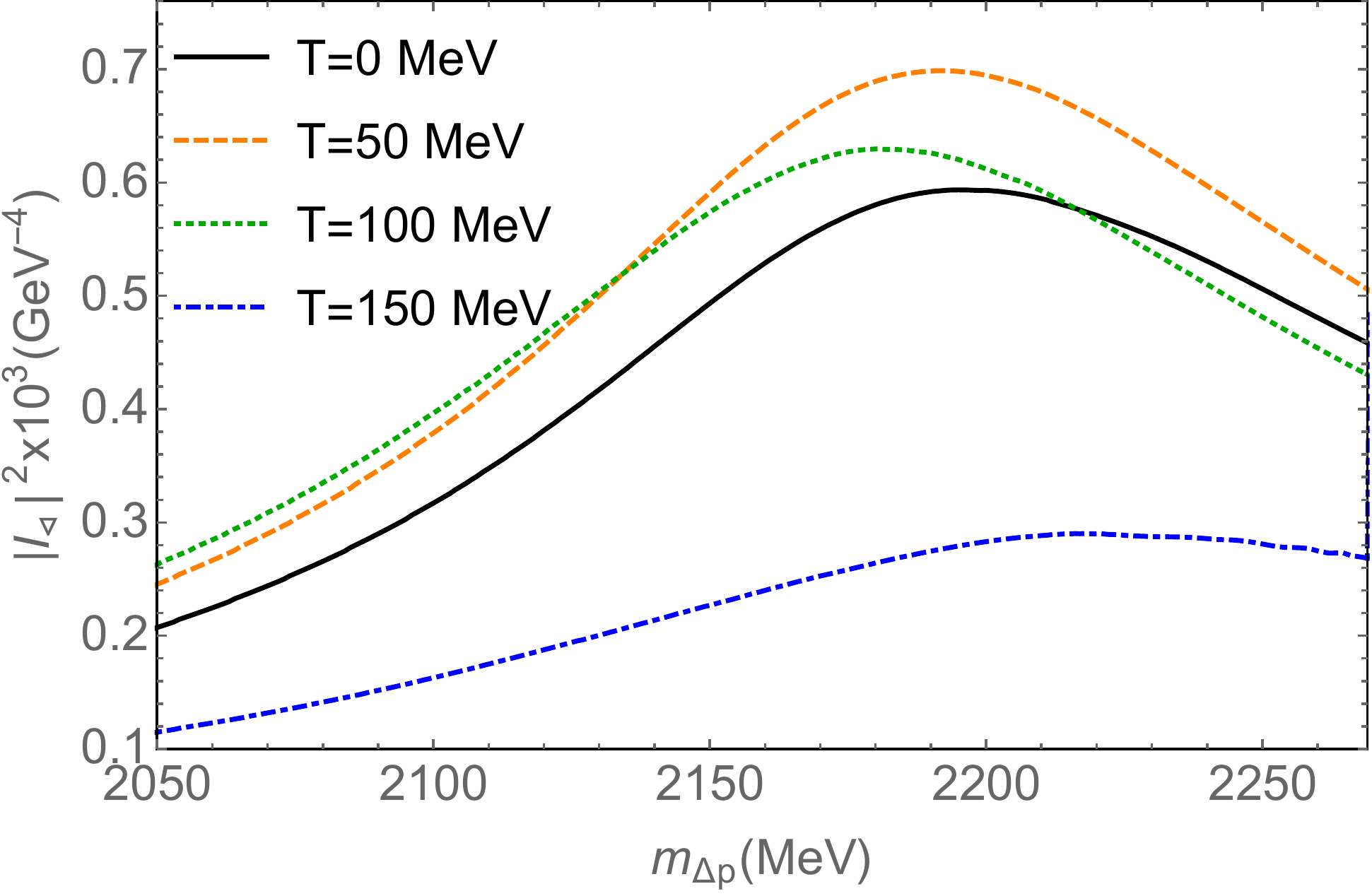}
\caption{\label{fig:deutspectrum} Spectrum of $\Delta p$ in the region corresponding to the presumed $d^*$ resonance.
If the triangle mechanism of figure~\ref{fig:triangle-DELTAPN} is behind the apparently resonant cross-section, 
a thermal bath will significantly damp it for temperatures above 100 MeV: the difference between the top (with vacuum particle properties) and bottom (with thermal properties) plots shows once more that it is very important to know the behavior of the particle poles in the medium, as they provide the dominant effect by erasing or fuzzying the kinematic coincidence that produces Landau singularities.
}
\end{figure}

What the plot shows is that the position of the triangle+threshold singularity is affected by the T-modified baryon parameters, shifting within the order of magnitude of that change, but it is not washed out. The mass-shift is sufficiently intense that a detailed calculation with an evolution code might serve a prediction good enough to test against other interpretations.
If we turn to the $\Delta p$ spectrum where the excited deuteron candidate decaying to $d\pi$ should appear, see figure~\ref{fig:deutspectrum}, we find, in accordance with the intervening triangle diagram in figure~\ref{fig:triangle-DELTAPN}, 
that the leading thermal effect is found (bottom plot) upon taking into account the thermal masses and widths of the particles involved; and most importantly, that the resonant-like behavior of this spectrum, that suggests the possibility of an excited deuteron state, is suppressed at the highest temperatures that we considered, around 150 MeV. 

\section{Flight time of the intermediate state}\label{sec:flight}
The first condition for a triangle singularity of the Feynman diagram is that all particles $i=1,2,3$ in the loop satisfy on-shell kinematics (with velocity $\beta_i$ in the rest frame of $A$). The second is that particle 3 (emitted after the decay of particle 1) catches up with particle 2 in its classical motion. This gives a characteristic time scale for the triangle process. If this time is extremely longer than the lifetime of the RHIC fireball, most of the propagation of the intermediate state will take place outside the medium, so that its effect might be less important, or at least require a more sophisticated treatment taking into account the nonequilibrium dynamics up to freeze out. 
To assess the applicability of our thermal calculations (that are set up for infinite matter) in real heavy ion collisions requires an estimate of the flight time, that we here provide.

In the rest frame of $A$, particle 2 travels a distance $x$ in a time $\tau_A$ since the decay of the originating particle $A$:
\begin{equation} \label{x}
x = \tau_A \beta_2\ .
\end{equation}
Particle 3 is emitted after a typical delay given by the propagation of particle 1, inversely proportional to its decay width $\tau_1 \simeq \frac{\gamma(\beta_1)}{\Gamma_1}$ (the intrinsic width is diminished by the dilation $\gamma$ factor due to the motion of particle 1).

The motion of particle 1 followed by the later displacement of particle 3, always in the rest frame of $A$,
yields
\begin{equation}
\sqrt{x_1^2+\Delta y^2} + \sqrt{x_3^2 + \Delta y^2} = \tau_1 \beta_1 + (\tau_A-\tau_1)\beta_3 \ .
\end{equation}
For an order of magnitude estimate, the transverse displacement $\Delta y$ can be ignored (the kinematics is quite collinear, the more so the larger $\beta_2$). Thus, the condition for particles 2 and 3 to meet on-shell and complete the singular triangle requires that particle 3 reaches $x$ at $\tau_A$, that is, $x=x_1+x_3$ implying 
\begin{equation} \label{x2}
 x = \tau_1 \beta_1 + (\tau_A-\tau_1)\beta_3 \ .
\end{equation}
Therefore, combining Eq.~(\ref{x}) and~(\ref{x2}),
\begin{equation}
\tau_A \beta_2 =  \left( \tau_A - \frac{\gamma(\beta_1)}{\Gamma_1} \right)\beta_3  + \frac{\beta_1\gamma(\beta_1)}{\Gamma_1}
\end{equation}
we can obtain 
\begin{equation}\label{flighttime}
\boxed{\tau_A =\frac{\gamma(\beta_1)}{\Gamma_1} \frac{\beta_3-\beta_1}{\beta_3-\beta_2}} \ .
\end{equation}

To employ Eq.~(\ref{flighttime}) we need to know the three velocities.
$\beta_1= q/\sqrt{q^2+m_1^2}$ can be obtained from the decay vertex of particle 1, taking into account that
\begin{equation}
q = \lambda^{1/2}(m_A^2,m_1^2,m_2^2)/(2m_A)\ .
\end{equation}
As for $\beta_2$ and $\beta_3$, they are expressed~\cite{Guo:2019twa} in terms of $\beta=k/E_c$, that of the 23 subsystem in the rest frame of $A$, by 
\begin{eqnarray}
\beta_2 = \beta \frac{E_2^*-p_2^*/\beta}{E_2^*-p_2^* \beta} \\
\beta_3 = \beta \frac{E_3^*+p_2^*/\beta}{E_3^*+p_2^* \beta} 
\end{eqnarray}
that show how $\beta_3>\beta>\beta_2$ so that particle 3 can recover its initial delay and reach particle 2.

The starred quantities are in the rest frame of the 23 system instead of the rest frame of $A$. They are given by 
\begin{eqnarray}
E_2^* =  \frac{m_C^2+m_2^2-m_3^2}{2m_C} \\
p_2^* = \lambda^{1/2}(m_C^2,m_2^2,m_3^2)/(2m_C)
\end{eqnarray}
and similarly for particle 3. 

The expressions leading to Eq.~(\ref{flighttime}) are easily coded, and we obtain the numerical evaluation for the flight times of the intermediate states listed in Table~\ref{tof}.

 \begin{center}
 \begin{table}[h]
\caption{Flight time and its inverse for the five intermediate states considered in the examples of section~\ref{Results}. (The  $\Delta^+ p n $ case is particularly sensitive to the value of $M_C$, so we give two different evaluations). Those cases with $\tau_A$ above tens of fm are unlikely to be well approximated by our infinite medium computation. 
\label{tof}
} 
 \begin{center}
\vskip1.5mm
\begin{tabular}{c|c|c}
\hline
\hline
Triangle diagram (value of $m_C$) & $\tau_A $ (fm/c)& $\tau_A^{-1}$ (MeV)  \\ 
\hline
$K^{\ast} \Sigma \pi $ ($m_C = 1400 $ MeV)   & 3.9    & 50.4 \\ 
$K^{*} \overline{K} K$ ($m_C = 988.4$ MeV)   & 10     & 19.7 \\
$D ^{*} D^{*} D$       ($m_C = 3871.71$ MeV) & 2346   & 0.084 \\
$D^* D D$              ($m_C = 3729.82$ MeV) & 3595   & 0.055 \\
$\Delta^+ p n $        ($m_C = 1877.84$ MeV) & 8.4    & 23.5 \\
$\Delta^+ p n $        ($m_C = 1880$ MeV)    & 1.25   & 157 \\
\hline
\hline
\end{tabular}
 \end{center}
 \end{table}
 \end{center}

\newpage

\section{Discussion}

Our example reactions in subsections~\ref{subsec:lambda} through \ref{subsec:dstar}
clearly establish that triangle (and threshold) singularities can be seriously affected by the equilibrated medium created in relativistic heavy ion collisions, and that their disappearance
due to the kinematic conditions for the triangle singularity to occur not being met, due to thermal mass shifts and widths, provides a new tool for hadron spectroscopy classification of final state enhancements.

We have provided calculations in an equilibrated medium at temperatures typical of the hadron phase, between 100 and 150 MeV, probably present in the last phases of relativistic heavy-ion collision experiments~\cite{Abelev:2013vea} at large multiplicity, showing how the triangle singularities are washed away (computations at 50 MeV, less relevant for high-energy collider experiments, are provided as a benchmark). In good logic, if these are the lowest temperatures reported in the final stages of the collisions, dearth of a certain kinematic singularity implies that it will not be present at all, as it will not be realized at any earlier stage of the collision with naturally higher $T$, provided the two following conditions are met. 

Although our computations are carried out for infinite matter, a first real-world condition for heavy-ion collider experiments is that the characteristic time scale of the triangle loop (controlled by the width  $\Gamma_1$) is short enough for the process to effectively take place during the lifetime of the fireball. This seems to exclude cases such as those in subsections~\ref{subsec:X1} and~\ref{subsec:X2}. The second condition is that $M_i$, $\Gamma_i$, $i=1,2,3$ acquire a sufficient temperature (or density) dependence so that the medium thwarts the kinematic coincidence yielding the singularity.

Awareness of this phenomenon in analyzing heavy-ion collision data may therefore prove useful. 
In fact, we are not aware of any heavy-ion collision sighting of the resonances or processes listed in table 1 of~\cite{Guo:2019twa} as being strongly influenced by the triangle mechanism. A dedicated search would be useful to decide which ones are indeed absent in that hot environment.

The existence of triangle singularities does not necessarily imply an observable increase of the width of the initial A system that may enjoy other decay paths (if it is a particle) or reaction paths (if a composite multiparticle system). The relevance of these singularities is that a subset of the decay products of A
may present a sharp enhancement at very specific values of, for example, $m_{C}$ (whereas perhaps in other reaction branches $A\to DE$ or $A\to FG$, $m_{D}$ or $m_{F}$ would feature no such enhancement). In consequence, it might appear to an experimental observer that a new particle or resonance with the quantum numbers of $C$ would have been discovered for example, $C=\pi^+\pi^-$ in Eq.~(\ref{reaccion2}). With modern detectors and software allowing the matching of momenta of particle pairs across complicated collisions, many potential discoveries are reported: interpreting them as hadron resonances requires care that no reaction enhancement due to other reasons is causing the relative excess in the data.
The effect of the medium, when active, is then to erase the coincidence that causes the data to blip at specific kinematics. While this can be drastic for this local enhancement, the overall global reaction properties of $A$ probably react more linearly to the presence of the medium.

We have not attempted to provide fully normalized cross-sections that depend on much more detailed physics of the collision~
\cite{Zhang:2020dwn}.

 Instead, we have only shown the relative spectral strength at different energies for the various processes. Since other reactions that affect the decaying state $A$ when {\it in medio} differently from when {\it in vacuo}, such as $AX_1\to X_2 X_3$, will not generally have a strong energy dependence peaking at the same kinematics of the Landau triangle singularity, their depleting the abundance of $A$ should not affect our considerations.

Our estimates indicate that the main effect behind the disappearance of the singularities is the modified {\rm in medio} mass and width of the participating hadrons. However, in the case of broader structures (see figure~\ref{fig:triangle-KSTARSIGMAPION}), the structure of the triangle itself is also affected by the thermal medium. Our method can be deployed as a starting point to study multiple such singularities in heavy ion collisions.

\section*{Acknowledgements}
We thank Fernando Silveira Navarra, \'Angel G\'omez-Nicola, Juan Torres Rinc\'on, Feng-Kun Guo and Eulogio Oset for discussion  on particular aspects of the investigation, and the last two for reading the draft and pointing out improvements.
\paragraph{Funding information}
Work supported by Brazilian funding agencies CNPq (contracts 308088/2017-4 \& 400546/2016-7) and FAPESB (contract INT0007/2016); Spain's MINECO FPA2016-75654-C2-1-P, MICINN PID2019-108655GB-I00 \& -106080GB-C21;  EU’s 824093 (STRONG-2020); and UCM's 910309 group grants \& IPARCOS.

\newpage

\end{document}